\documentclass{iau} 
\usepackage{graphicx}

\title[High dispersion spectroscopy of solar-type superflare stars] 
{High dispersion spectroscopy of solar-type superflare stars with Subaru/HDS 
\thanks{This study is based on observational data collected with Subaru Telescope, 
which is operated by the National Astronomical Observatory of Japan. }}

\author[Notsu, Y., Honda, S., Maehara, H., et al.]   
{Yuta Notsu$^{1,*}$,
Satoshi Honda$^2$,
Hiroyuki Maehara$^3$,
Shota Notsu$^1$,
Takuya Shibayama$^4$,
Daisaku Nogami$^1$,
\and Kazunari Shibata$^5$}

\affiliation{
$^*$email: {\tt ynotsu@kwasan.kyoto-u.ac.jp} \\[\affilskip]
$^1$Department of Astronomy, Kyoto University, Kitashirakawa-Oiwake-cho, Sakyo-ku, Kyoto, Japan, 606-8502 \\
$^2$Center for Astronomy, University of Hyogo, 407-2, Nishigaichi, Sayo-cho, Sayo, Hyogo, Japan, 679-5313 \\
$^3$Okayama Astrophysical Observatory, National Astronomical Observatory of Japan, 3037-5 Honjo, Kamogata, Asakuchi, Okayama, Japan, 719-0232 \\
$^4$Institute for Space-Earth Environmental Research, Nagoya University, Furo-cho, Chikusa-ku, Nagoya, Aichi, Japan, 464-8601\\
$^5$Kwasan and Hida Observatories, Kyoto University, Yamashina-ku, Kyoto, Japan, 607-8471
}

\pubyear{2015}
\volume{320}  
\jname{Solar and Stellar Flares and Their Effects on Planets}
\editors{A.G. Kosovichev, S.L. Hawley \& P. Heinzel, eds.}
\begin{document}

\maketitle

\begin{abstract}
\\
\ \ \
We carried out spectroscopic observations with Subaru/HDS of 50 solar-type superflare stars found from Kepler data.
More than half (34 stars) of the target stars show no evidence of the binary system, 
and we confirmed atmospheric parameters of these stars are roughly 
in the range of solar-type stars.
\\ 
\ \ \
We then conducted the detailed analyses for these 34 stars. 
First, the value of the ``$v\sin i$" (projected rotational velocity) measured from spectroscopic results 
is consistent with the rotational velocity estimated from the brightness variation. 
Second, there is a correlation between the amplitude of the brightness variation and the intensity of Ca II IR triplet line. 
All the targets expected to have large starspots because of their large amplitude of the brightness variation 
show high chromospheric activities compared with the Sun. 
These results support that the brightness variation of superflare stars
is explained by the rotation of a star with large starspots.
\end{abstract}

\firstsection 
\section{Introduction}\label{sec:intro}
\noindent
\ \ \
Flares are energetic explosions in the stellar atmosphere,
and are thought to occur by intense releases of magnetic energy stored around starspots,  
like solar flares (e.g., \cite[Shibata \& Magara 2011]{Shibata2011}).
Superflares are flares 10$\sim$10$^{6}$ times more energetic 
($\sim$10$^{33-38}$erg; \cite[Schaefer et al. 2000]{Schaefer2000}) than the largest solar flares ($\sim$10$^{32}$erg).
Recently, we analyzed the data of the Kepler space telescope (\cite[Koch et al. 2010]{Koch2010}), 
and discovered more than 1000 superflares on a few hundred solar-type (G-type main-sequence) stars 
(\cite[Maehara et al. 2012]{Maehara2012}, \cite[2015]{Maehara2015}; 
\cite[Shibayama et al. 2013]{Shibayama2013}; \cite[Candelaresi et al. 2014]{Candelaresi2014}).
We here define solar-type stars as the stars that have 
a surface temperature of $5100\leq T_{\rm{eff}}\leq 6000$K and a surface gravity of $\log g \geq 4.0$.
\\
\ \ \
With these data, we studied the statistical
properties of the occurrence rate of superflares,
and found that the occurrence rate ($dN/dE$) of the superflare versus the flare energy ($E$) has a power-law distribution
of $dN/dE\propto E^{-\alpha}$, where $\alpha\sim 2$ 
(\cite[Maehara et al. 2012]{Maehara2012}, \cite[2015]{Maehara2015}; \cite[Shibayama et al. 2013]{Shibayama2013}), 
and this distribution is roughly similar to that for the solar flare.
\\
\ \ \
Many of the superflare stars show quasi-periodic brightness variations with a typical period of from one day to a few tens of days.
The amplitude of these brightness variations is in the range of 0.1-10\% (\cite[Maehara et al. 2012]{Maehara2012}), 
and is much larger than that of the solar brightness variation (0.01-0.1\%) caused by the existence of sunspots on the rotating solar surface.
\cite{YNotsu2013} showed that the above brightness variations of superflare stars can be well explained 
by the rotation of a star with fairly large starspots, taking into account the effects of inclination angle and the spot latitude.
\\ 
\ \ \
\cite{YNotsu2013} compared the superflare energy and frequency with the rotation period, 
assuming that the brightness variation corresponds to the rotation. 
They then found slowly rotating stars can still produce as energetic flares as those of more rapidly rotating stars, 
though the average flare frequency is lower for more slowly rotating stars.
\cite{YNotsu2013} also clarified that the superflare energy is related to the total coverage of the starspots, 
and that the energy of superflares can be explained by the magnetic energy stored around these large starspots.
In addition, \cite{Shibata2013} suggested, on the basis of theoretical estimates, that the Sun can generate large magnetic flux sufficient for causing
superflares with an energy of $10^{34}$ erg within one solar cycle.
\\ 
\ \ \ 
The results described above are, however, only based on Kepler monochromatic photometric data. 
We need to spectroscopically investigate whether these brightness variations are explained by the rotation, and whether superflare stars have large starspots.
The stellar parameters and the binarity of the superflare stars are also needed to be investigated with spectroscopic observations
in order to discuss whether the Sun can really generate superflares.
We have then performed high-dispersion spectroscopy of superflares stars (50 stars in total).
We describe the results of this observation in the following.
\vspace{-3.5mm}
\section{Observations}\label{sec:obs}
\noindent
\ \ \
In this observation, we observed 50 solar-type superflare stars 
by using High Dispersion Spectrograph (HDS; \cite[Noguchi et al. 2002]{Noguchi2002}) at the 8.2m Subaru telescope 
on 6 nights during 2011$\sim$2013 (\cite[Notsu et al. 2013a]{SNotsu2013}, \cite[2015a]{YNotsu2015a} \& \cite[2015b]{YNotsu2015b}; 
\cite[Nogami et al. 2014]{Nogami2014}; \cite[Honda et al. 2015]{Honda2015}).
These 50 target stars were selected from superflare stars reported by our previous researches (e.g., \cite[Shibayama et al. 2013]{Shibayama2013}).
Spectroscopic resolution ($R=\lambda/\Delta\lambda$) of each observation date is $R=50,000\sim100,000$.
The wavelength range is 6100$\sim$8800\AA.
This range includes Ca II near-IR triplet (8498/8542/8662\AA) and H$\alpha$ (6563\AA), 
which are well-known indicators of stellar chromospheric activity. 
More details of the observations and the target stars are described in \cite{YNotsu2015a}.
\vspace{-3.5mm}
\section{Results and discussion}\label{sec:results-dis}
\subsection{Stellar parameters}\label{subsec:atmos}
\noindent
\ \ \
As a result of the observations, 
we found more than half (34 stars) of our 50 targets have no evidence of binary system (\cite[Notsu et al. 2015a]{YNotsu2015a}). 
Among the remaining 16 stars, 12 stars show double line profiles or radial velocity shifts, which are expected be caused by 
the orbital motion of binary system. 
The other 4 stars have visual companion stars. 
\\
\ \ \
We then estimated effective temperature ($T_{\rm{eff}}$), surface gravity ($\log g$), metallicity ([Fe/H]),
and projected rotation velocity ($v\sin i$)
of these 34 ``single" superflare stars on the basis of our spectroscopic data (\cite[Notsu et al. 2015a]{YNotsu2015a}). 
We here estimated $v\sin i$ by measuring Doppler broadening of photospheric lines 
taking into account macroturbulence and insturumental broadening 
(cf. \cite[Takeda et al. 2008]{Takeda2008}; \cite[Notsu et al. 2015a]{YNotsu2015a}). 
\\
\ \ \
We confirmed that stellar atmospheric parameters ($T_{\rm{eff}}$, $\log g$, and [Fe/H]) of 
the 34 target stars are roughly in the range of ordinary solar-type (G-type main sequence) stars.
In particular, the temperature, surface gravity, and brightness variation period ($P$) 
of 9 stars are in the range of ``Sun-like" stars ($5600\leq T_{\rm{eff}}\leq 6000$K, $\log g\geq$4.0, and $P>$10 days).
Five of the 34 target stars are fast rotators ($v \sin i \geq 10$km s$^{-1}$), 
while 22 stars have relatively low $v \sin i$ values ($v \sin i<5$km s$^{-1}$). 
These results suggest that stars whose spectroscopic properties similar to the Sun can have superflares, 
and this supports the hypothesis that the Sun might cause a superflare.
In the following, we conducted the detailed analyses for these 34 single stars (\cite[Notsu et al. 2015b]{YNotsu2015b}).
\subsection{Rotational velocity and inclination angle}\label{subsec:vsini}
\noindent
\ \ \
We here compare ``$v\sin i$" with the brightness variation period ($P$), 
and consider whether the brightness variation of superflare stars is explained by the rotation.
Assuming that the brightness variations are caused by the rotation
of the stars with starspots, we can estimate the rotational velocity ($v_{\rm{lc}}$) 
from $P$ and $R_{\rm{s}}$ (stellar radius) by using 
\begin{equation}\label{eq:vlc}
v_{\rm{lc}}=\frac{2\pi R_{\rm{s}}}{P} \ .
\end{equation}
\begin{figure}[htbp]
\begin{center}
 \includegraphics[width=0.55\textwidth]{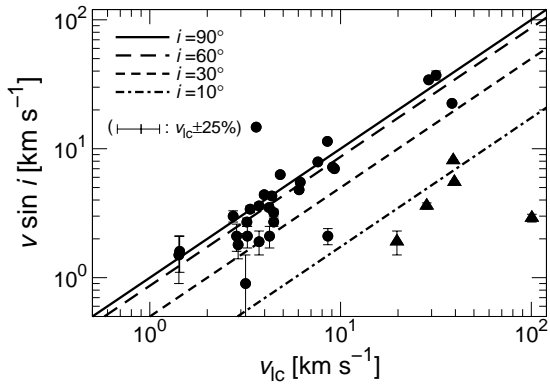}
 \caption{
Projected rotational velocity ($v \sin i$) as a function of the stellar rotational velocity ($v_{\rm{lc}}$) estimated from 
the period of the brightness variation and stellar radius. 
The typical error of $v_{\rm{lc}}$ is about $\pm$25\%~of each value, considering errors of $P$ and $R_{\rm{s}}$ (\cite[Notsu et al. 2015a]{YNotsu2015a}).
The solid line represents the case that our line of sight is vertical to the stellar rotation axis ($i=90^{\circ}$; $v \sin i=v_{\rm{lc}}$).
We also plot three different lines, which correspond to smaller inclination angles ($i=60^{\circ}$, $30^{\circ}$, $10^{\circ}$).
Filled triangles represent superflare stars whose inclination angle is especially small ($i\leq 13^{\circ}$), 
while filled circles represent the other stars ($i>13^{\circ}$).
}
\label{fig:vlc-vsini}
\end{center}
\end{figure}
\\
\ \ \
In Figure \ref{fig:vlc-vsini}, we plot $v \sin i$ as a function of the $v_{\rm{lc}}$. 
Some data points in Figure \ref{fig:vlc-vsini} show differences between the values of $v_{\rm{lc}}$ and $v \sin i$.
The projected rotational velocity ($v \sin i$) tends to be smaller than $v_{\rm{lc}}$.
Such differences should be explained by the inclination effect, as in \cite[Notsu et al. (2013a)]{SNotsu2013}.
On the basis of $v \sin i$ and $v_{\rm{lc}}$, the stellar inclination angle ($i$) can be estimated by using the following relation:
\begin{equation}\label{eq:inc}
i=\arcsin\Biggl(\frac{v \sin i}{v_{\rm{lc}}}\Biggr) \ .
\end{equation}
\ \ \
In Figure \ref{fig:vlc-vsini}, we also show four lines indicating $i=90^{\circ}$ ($v \sin i=v_{\rm{lc}}$), $i=60^{\circ}$, $i=30^{\circ}$, and $i=10^{\circ}$.
This figure shows two following important results. 
First, for almost all the stars (33 stars), the relation ``$v \sin i\lesssim v_{\rm{lc}}$" is satisfied.
This is consistent with our assumption that the brightness variation is caused by the rotation 
since the inclination effect mentioned above can cause the relation ``$v \sin i\lesssim v_{\rm{lc}}$" if 
$v_{\rm{lc}}$ values really correspond to the rotational velocities (i.e. $v=v_{\rm{lc}}$).
This is also supported by another fact that the distribution of the data points in Figure \ref{fig:vlc-vsini} are not random. 
Their distribution is expected to be much more random if the brightness variations have no relations with the stellar rotation.
Second, stars that are distributed in the lower right side of Figure \ref{fig:vlc-vsini} are expected to have small inclination angles and to be nearly pole-on stars.
In this figure, we distinguish such five stars with especially small inclination angle ($i\leq13^{\circ}$) 
from the other stars, using filled triangle data points. 
\begin{figure}[htbp]
 \begin{center}
 \includegraphics[width=0.6\textwidth]{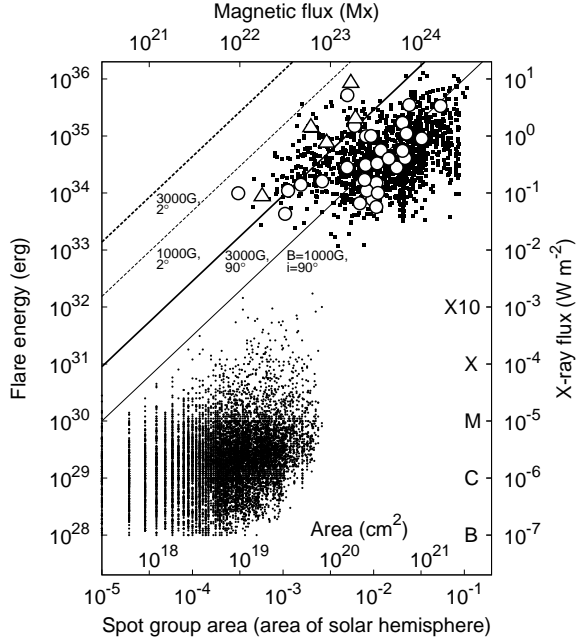}
\caption{Scatter plot of the flare energy as a function of the spot coverage.
The data of superflares on solar-type stars (filled squares) and solar flares (filled circles) in this figure are 
completely the same as those in Figure 10 of \cite{YNotsu2013}.
Thick and thin solid lines corresponds to the analytic relation 
between the spot coverage and the flare energy, which is obtained from Equation (14) of \cite{YNotsu2013} for $B$=3,000G and 1,000G. 
The thick and thin dashed lines correspond to the same relation in case of $i=2^{\circ}$ (nearly pole-on) for $B$=3,000G and 1,000G. 
Open circle and triangle points on the filled squares represent the data points of the most energetic superflare event reported in \cite{Shibayama2013}
of the 34 target superflare stars. 
Open triangles represent superflare stars whose inclination angle is especially small ($i\leq 13^{\circ}$), 
while open circles represent the others ($i>13^{\circ}$). This classification is on the basis of Figure \ref{fig:vlc-vsini}.}
\label{fig:spotene}
\end{center}
\end{figure}
\\
\ \ \ 
We can confirm the above inclination effects from another point of view.
Figure \ref{fig:spotene} is a scatter plot of the flare energy of superflares and solar flares as a function of the spot coverage.
The spot coverage of superflare stars is calculated from the amplitude of stellar brightness variations.
Thick and thin solid lines correspond to the analytic relation 
between the spot coverage and the flare energy, which is obtained from Equation (14) of \cite{YNotsu2013} in case of $i=90^{\circ}$ for $B$=3,000G and 1,000G. 
The thick and thin dashed lines correspond to the same relation in case of $i=2^{\circ}$ (nearly pole-on) for $B$=3,000G and 1,000G, 
assuming that the brightness variation becomes small as a result of the inclination effect.
These lines are considered to give an upper limit of superflare energy for each inclination angle.
Considering these things, the superflare stars located in the upper left side of this figure are expected to have a low inclination angle.
Open triangles in Figure \ref{fig:spotene} correspond to superflare stars whose inclination angle is especially small ($i\leq 13^{\circ}$)
 on the basis of Figure \ref{fig:vlc-vsini}, 
while open circles represent the other stars ($i>13^{\circ}$). This classification is the same as that in Figure \ref{fig:vlc-vsini}.
All of the five triangle data points are located above the thick solid line ($i=90^{\circ}$ and $B$=3000G). 
This means that these stars are confirmed to have low inclination angle on the basis of both of the Figures \ref{fig:vlc-vsini} and \ref{fig:spotene}.
As a result, these two figures (Figures \ref{fig:vlc-vsini} and \ref{fig:spotene}) are confirmed to be consistent. 
In other words, the stellar projected rotational velocity spectroscopically measured is consistent with the rotational velocity estimated from the brightness variation.
This fact supports that the brightness variation of superflare stars is caused by the rotation.
\subsection{Stellar chromospheric activity and starspots of superflare stars}\label{subsec:CaII-spot}
\noindent
\ \ \
In order to investigate the chromospheric activity of the target stars, 
we measured $r_{0}$(8498), $r_{0}$(8542), $r_{0}$(8662), and $r_{0}$(H$\alpha$) index, 
which are the residual core flux normalized by the continuum level 
at the line cores of the Ca II IRT and H$\alpha$, respectively.
These indexes are known to be good indicators of stellar chromospheric activity (e.g., \cite[Takeda et al. 2010]{Takeda2010}; \cite[Notsu et al. 2013a]{SNotsu2013}).
As the chromospheric activity is enhanced, the intensity of these indicators becomes large 
since a greater amount of emission from the chromosphere fills in the core of the lines. 
\begin{figure}[htbp]
\begin{center}
 \includegraphics[width=0.48\textwidth]{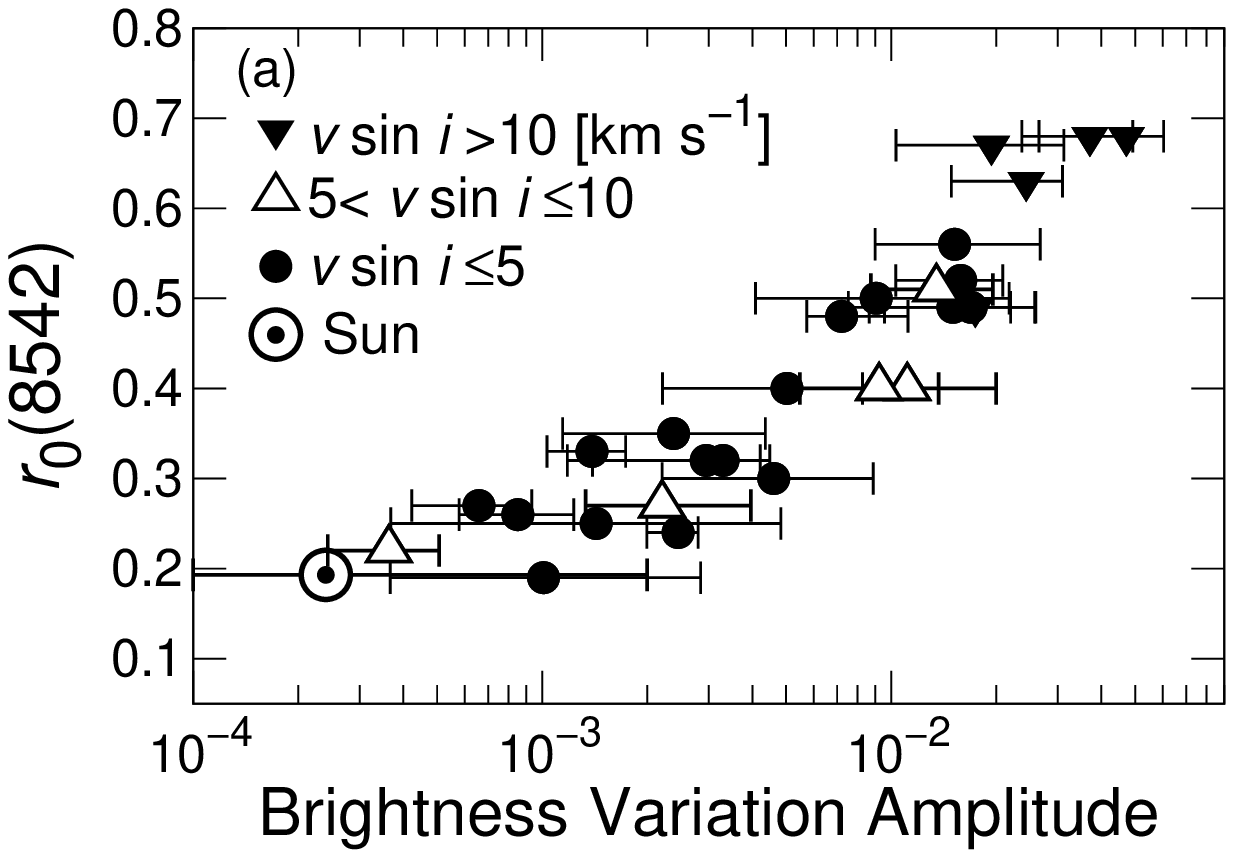} 
\hspace{0.01\textwidth}
 \includegraphics[width=0.48\textwidth]{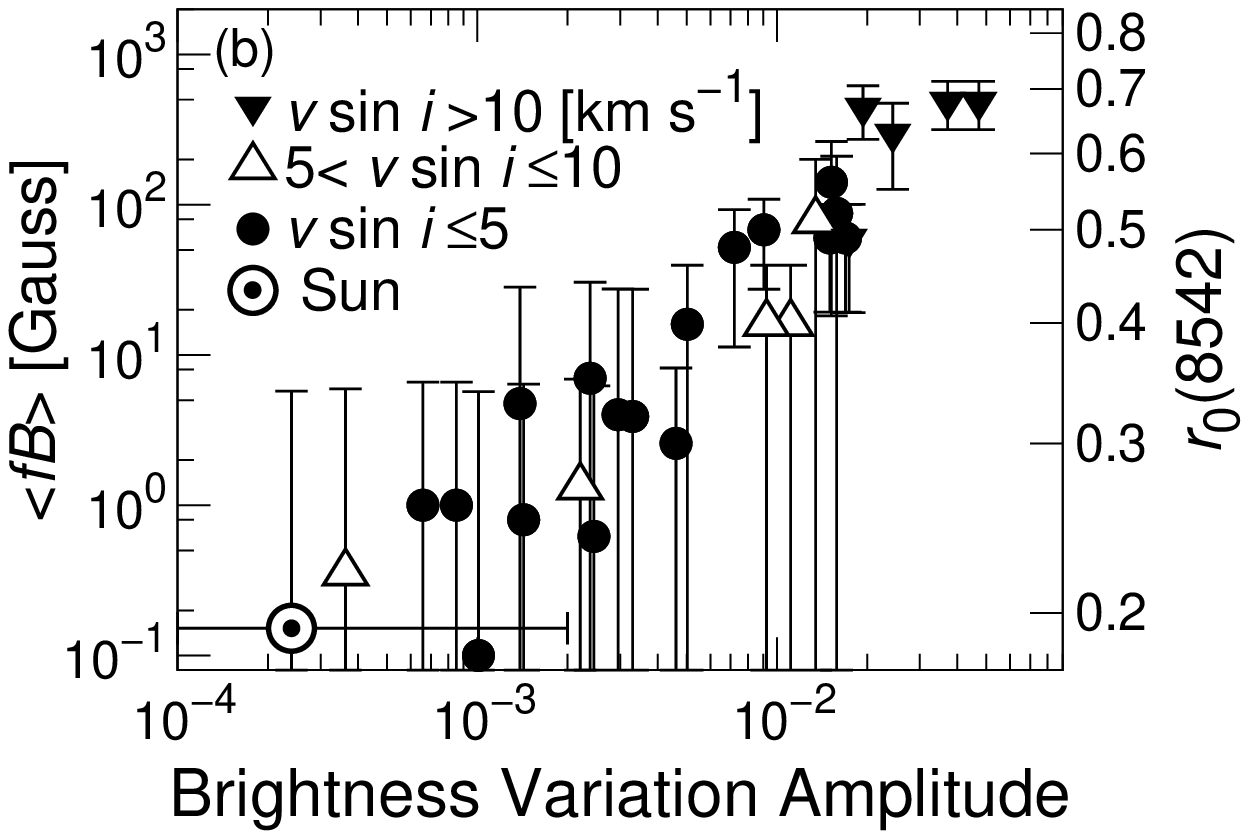} 
 \caption{
(a) $r_{0}$(8542) as a function of
the amplitude of stellar brightness variation ($\langle$BVAmp$\rangle$).
The results of the target superflare stars are plotted, being classified into three groups 
on the basis of $v\sin i$ (projected rotation velocity).
The solar value is plotted by using a circled dot point.
(b) $\langle fB\rangle$ as a function of amplitude of stellar brightness variation.
$\langle fB\rangle$ values are estimated from $r_{0}(8542)$
on the basis of the rough relation estimated in \cite{YNotsu2015b}.}
\label{fig:ampr0fB}
\end{center}
\end{figure}
\\
\ \ \
The $r_{0}$(8542) values are plotted in Figure \ref{fig:ampr0fB} (a) 
as a function of the amplitude of stellar brightness variation ($\langle$BVAmp$\rangle$) of Kepler data.
First, almost all the target superflare stars are more active compared with the Sun from the viewpoint of the $r_{0}$(8542) index.
In other words, the mean magnetic field strength of the target stars can be higher than that of the Sun.
Next, in this figure, there is a rough positive correlation between $r_{0}$(8542) and $\langle$BVAmp$\rangle$.
Assuming that the brightness variation of superflare stars is caused by the rotation of a star with starspots, 
the brightness variation amplitude ($\langle$BVAmp$\rangle$) corresponds to the starspot coverage of these stars. 
Then, we can say that there is a rough positive correlation between the starspot coverage and chromospheric activity level ($r_{0}$(8542)).
This rough correlation shows us that all the target stars expected to have large starspots 
on the basis of their large amplitude of the brightness variation show high magnetic activity compared with the Sun.
In other words, our assumption that the amplitude of the brightness variation correspond to the spot coverage is supported, 
since high magnetic activity, which are confirmed by using $r_{0}$(8542) values, are considered to be caused by the existence of large starspots.
\\ 
\ \ \ 
In Figure \ref{fig:ampr0fB} (b), we also plot $\langle fB\rangle$ values as a function of $\langle$BVAmp$\rangle$.
$\langle fB\rangle$ values are estimated from $r_{0}(8542)$
on the basis of the rough relation estimated in \cite{YNotsu2015b} with spectroheliographic observation of a solar active region.
With this figure, we can see the same conclusion as we did with Figure \ref{fig:ampr0fB} (a), 
though the errors of $\langle fB\rangle$ values are a bit large especially for less active stars.
We also investigated the emission flux of Ca II IRT and H$\alpha$ lines for reference in \cite{YNotsu2015b}, 
and confirmed basically the same conclusions as we did here with $r_{0}$(8542) index. 
\vspace{-3.5mm}
\section{Summary}\label{sec:summary}
\ \ \
Superflares are very large flares that release total energy 10$\sim$10$^4$ times greater than 
that of the biggest solar flares ($\sim$10$^{32}$ erg). 
Recent Kepler-space-telescope observations found more than 1000 superflares 
on a few hundred solar-type stars. 
Such superflare stars show quasi-periodic brightness variations with the typical period of from one to a few tens of days. 
Such variations are thought to be caused by the rotation of a star with large starspots. 
However, spectroscopic observations are needed in order to confirm whether the variation is 
really due to the rotation and whether superflares can occur on ordinary single stars similar to our Sun.
\\
\ \ \
We have carried out spectroscopic observations of 50 solar-type superflare stars with Subaru/HDS. 
As a result, more than half (34 stars) of the target stars show no evidence of the binary system, 
and we confirmed stellar atmospheric parameters of these stars are roughly 
in the range of solar-type stars on the basis of our spectroscopic data.
\\
\ \ \
We then conducted the detailed analyses for these 34 stars. 
First, the value of the $v\sin i$ measured from spectroscopic results 
is consistent with the rotational velocity estimated from the brightness variation. 
Next, we measured the intensity of Ca II IRT and H$\alpha$ lines, which are good indicators of stellar chromospheric activity. 
The intensity of these lines indicates that the mean magnetic field strength ($\langle fB\rangle$) 
of the target superflare stars can be higher than that of the Sun.
We found a correlation between the amplitude of the brightness variation and the intensity of Ca II IRT. 
All the targets expected to have large starspots because of their large amplitude of the brightness variation show high 
chromospheric activity compared with the Sun.
These results support that the brightness variation of superflare stars is explained by the rotation with large starspots.

\end{document}